# Effect of 3d Metal (Co and Ni) Doping on the Superconductivity of $FeSe_{0.5}Te_{0.5}$

Anuj Kumar[1,2], R. P. Tandon[2], and V. P. S. Awana[1]

[1]Quantum Phenomena and Application Division, National Physical Laboratory (CSIR)
Dr. K. S. Krishnan Road, New Delhi-110012 INDIA
[2]Department of Physics and Astrophysics, University of Delhi, North Campus, New Delhi-110007 INDIA

We report the effect of 3d metal Cobalt (Co) and Nickel (Ni) doping on the $FeTe_{0.5}Se_{0.5}$ superconductor with the nominal composition range $Fe_{1-x}M_xTe_{0.5}Se_{0.5}$ (M = Co, Ni and x = 0.00, 0.01 0.02, 0.05, and 0.10). Samples are synthesized through standard solid state reaction route and all are crystallize in single phase tetragonal structure with space group *P4nmm*. The lattice parameters '*a*', '*c*' and volume decrease with increase in Co and Ni content, although not monotonically. In fact the '*a*' lattice parameter of Co doped samples is nearly unaffected for Co doped samples. The superconducting transition temperature ($T_c$) is measured from both DC and AC magnetic susceptibility, which decrease with increase in Ni or Co content. Both Co and Ni suppress $T_c$ and drive the system into the normal state. Interestingly, Ni suppresses the superconductivity much faster than the Co. This indicates less effect on in-plane Fe-Se distances and thus reduced disorder in case of Co when compared with Ni substation at Fe site in $FeTe_{0.5}Se_{0.5}$ superconductor. It is concluded that in-plane disorder on superconducting Fe-Se and Fe-Se-Se planes directly affects superconductivity in $FeTe_{0.5}Se_{0.5}$ superconductor.

*Index Terms*-$FeSe_{0.5}Te_{0.5}$, Fe chalcogenides, magnetization, plane disorder.

## I. INTRODUCTION

THE discovery of superconductivity in Fe-based pnictides compounds, $REO_{1-x}F_xFeAs$ was a real surprise and had attracted a lot of attention in condensed matter community over the last four years [1]-[5]. Due to the presence of Fe ion there has been particular interest in the possible connection between magnetism and superconductivity. They all commonly contain FeAs superconducting layers in the crystal structure. One of the silent features of FeAs based superconductor is their structural transition from a tetragonal to an orthorhombic phase at around 150 K [4] closely followed by the formation of spin density wave (SDW) ground state. The SDW is suppressed by electron/hole doping and superconductivity is observed. The FeAs based superconductors are the only known compounds other than the high temperature superconducting cuprates (HTSc) with their $T_c$ close to 56 K [3]. Now it is quite natural to think weather other Fe-based planer compounds exists that show superconductivity. In July 2008, Hsu et. al., reported superconductivity at 8 K in anti-PbO type FeSe compound [6]. The superconductivity is reported to increase to 27 K under modest pressure [7]. The crystal structure of FeSe is the simplest among Fe-based superconductors. Density functional calculations on FeS, FeSe and FeTe indicated that the strength of spin density wave (SDW) in FeTe and the possibility of higher $T_c$ in doped FeTe alloy compared to FeSe [8]. The enhancement of $T_c$ in Te substituted FeSe may be explained by the results of the density functional calculations. The superconducting state exists over quite a wide range of $T_c$ doping in the Fe (Se, Te) system (up to 90% Te substitution for Se in polycrystalline samples) with a maximum $T_c$ of 15 K [9], [10]. However, pure FeTe is not superconducting and the two end compounds FeSe and FeTe are structurally isomorphic but revels different physical and magnetic properties. FeSe has been studied quite extensively [11], a key observation is that the phase pure superconducting sample exists only for those samples prepared with intentional Se deficiency. FeSe comes in several phases: (a) a tetragonal phase α-FeSe with PbO-structure, (b) a NiAs-type β-phase with a wide range of homogeneity showing a transformation from hexagonal to monoclinic symmetry, and (c) a $FeSe_2$ phase that has the orthorhombic marcasite structure. The most studied of these compounds are the hexagonal $Fe_7Se_8$ phase, which is a ferrimagnet with Curie temperature at around 250 K [12], [13] and monoclinic $Fe_3Se_4$. Iron chalcogenides are low career density metals with high density of states where the electronic structure near the Fermi level is mainly influenced by Fe derived bands. Se derived bands lie well below the Fermi level and Fe-Se hybridization is week as compared to Fe-Fe interaction [8].

The $FeSe_{1-x}Te_x$ system has been intensively studied with respect to the interplay between structural of magnetic point of view and superconductivity also. In $FeSe_{1-x}Te_x$, Fe layers have a square lattice structure: however the positions of the Se/Te atoms above and below those planes break the transitional symmetry. Thus it is crystallographically appropriate to choose a unit cell with two Fe atoms per layer. The $FeSe_{1-\delta}$, $FeTe_{1-\delta}$ and $FeTe_{1-x}Se_x$ superconductors also show structural and magnetic transitions. It is still a question of debate whether the presence of excess interstitial Fe magnetic ions are responsible for such magnetic transitions [14]. So there are two position of Fe atom in $FeSe_{0.5}Te_{0.5}$ superconductor [15]. It is reported [16] that magnetic fluctuations play an important role and calculations showed a strong sensitivity of the magnetic moment on the so called "chalcogen height" i.e., the distance of the Se/Te atoms from the plane of the Fe atom. As the magnetic nature of Fe atom, therefore it would be interesting to substitute other transition metal at Fe site in FeSe$_{0.5}$Te$_{0.5}$ superconductor. The relationship between the external pressure effect and chemical pressure effect is unclear. To elucidate the mechanism for changes T$_c$ we investigated the effect of Cobalt (Co) and Nickel (Ni) substitution at Fe site in FeSe$_{0.5}$Te$_{0.5}$ superconductor and explore its structural, physical and magnetic properties. There are very few studies on such substitution have been reported in Fe$_{1.01-x}$Cu$_x$Se [17] system, and there is only one report for FeSe$_{0.5}$Te$_{0.5}$ [18] by Ni and Co at Fe site. Here one thing is important Co and Ni substitutions in un-doped Fe based oxypnictides induce superconductivity [19], [20] by providing charge carriers in the FeAs plane but the superconducting transition is low compared to fluorine doped samples. While here in FeSe and FeSe$_{0.5}$Te$_{0.5}$ samples there is a negative impact of Co and Ni on T$_c$, it means Fe is a good candidate to obtain superconductivity on chalcogenides and oxypnictides superconductors.

## II. EXPERIMENTAL

Two series of samples with composition Fe$_{1-x}$Co$_x$Se$_{0.5}$Te$_{0.5}$ (x = 0.0, 0.01, 0.02, 0.05 & 0.10) and Fe$_{1-x}$Ni$_x$Se$_{0.5}$Te$_{0.5}$ (x = 0.0, 0.01, 0.02, 0.05 & 0.10) were synthesized through standard solid state reaction route. Initially stoichiometric quantities of highly pure (> 3N) Fe, Se, Te, Co and Ni were ground, pelletized and then encapsulated in an evacuated (vacuum > 10$^{-3}$ Torr) quartz tube. The starting mixtures were heated slowly with the rate of 1$^o$C/min. at 700$^o$C for 12 h after which they were reground, pelletized, encapsulated and reheated with the same heating rate (1$^o$C/min.) at 750$^o$C for 24 h and then slowly cooled down to room temperature over a span of 12 h. All grindings were performed in high purity argon (3N) filled glove-box with oxygen and humidity content less than 1 ppm. The sintered samples are hard and stable in air but kept in evacuated desiccators to protect them from moisture. X-ray diffraction (XRD) for all synthesized samples was performed at room temperature in the scattering angle range of 10$^o$-60$^o$ in equal 2θ steps of 0.02$^o$ using a Rigaku Diffractometer with CuK$_\alpha$ (λ = 1.54Å). Rietveld analysis of XRD patterns were performed using the *FullProf* program. Temperature dependent AC and DC magnetization were performed on a physical property measurement system (PPMS-14T) made by Quantum design USA.

## III. RESULTS AND DISCUSSION

Fig.1 shows the observed and fitted X-ray diffraction pattern for FeSe$_{0.5}$Te$_{0.5}$, Fe$_{0.98}$Co$_{0.02}$Se$_{0.5}$Te$_{0.5}$ and Fe$_{0.95}$Ni$_{0.05}$Se$_{0.5}$Te$_{0.5}$ samples. All samples are single phase. All samples are fitted in tetragonal lattice with space group *P4/nmm*. The lattice parameters are obtained from respective Rietveld analysis for all the XRD patterns. The best fits were obtained within a χ$^2$ range of less than 2. It was observed that lattice parameters and unit cell volume decreases continuously with the substitution of Ni and Co at Fe site in FeSe$_{0.5}$Te$_{0.5}$. Decrease in volume of lattice indicates successful substitution of smaller Co$^{2+}$ and Ni$^{2+}$ ions at bigger Fe$^{2+}$ in doped

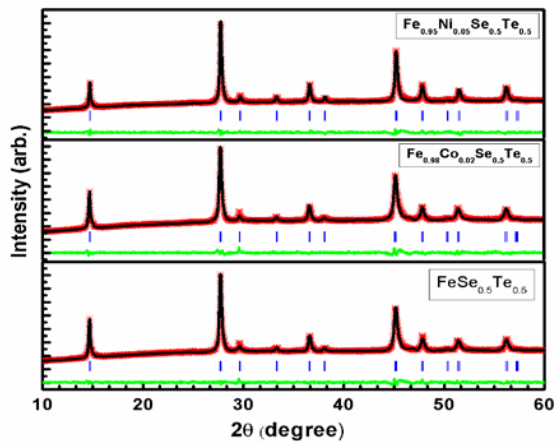

Fig. 1. Observed and fitted XRD patterns of FeSe$_{0.5}$Te$_{0.5}$, Fe$_{0.98}$Co$_{0.02}$Se$_{0.5}$Te$_{0.5}$ and Fe$_{0.95}$Ni$_{0.05}$Se$_{0.5}$Te$_{0.5}$ at room temperature.

FeSe$_{0.5}$Te$_{0.5}$ compound. The occupancy for the Fe was fixed to be 1 and after the refinement the obtained value of occupancy of Se is 0.40 and for Te is 0.60.

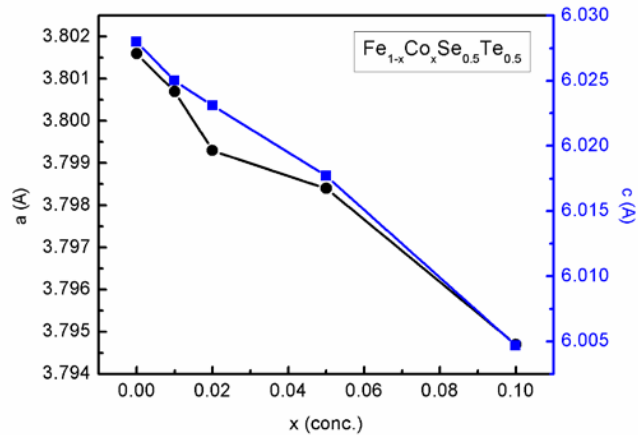

Fig. 2(a). Variation of lattice parameters '*a*' and '*c*' for Fe$_{1-x}$Co$_x$Se$_{0.5}$Te$_{0.5}$ as a function of dopant concentration x.

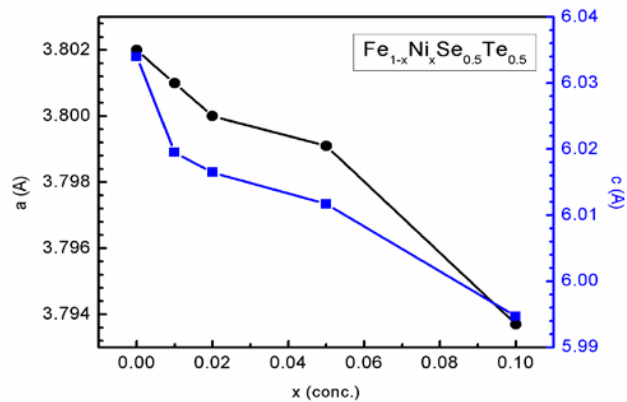

Fig. 2(a). Variation of lattice parameters 'a' and 'c' for $Fe_{1-x}Co_xSe_{0.5}Te_{0.5}$ as a function of dopant concentration x.

Fig. 2(a) and 2(b) shows the variation of 'a' and 'c' lattice parameters with Co and Ni doping in $Fe_{1-x}Co_xSe_{0.5}Te_{0.5}$ and $Fe_{1-x}Ni_xSe_{0.5}Te_{0.5}$ respectively. Both 'a' and 'c' decreases with increasing Co and Ni concentration, according to Vegard's law indicating the substitution of smaller cations compared to $Fe^{2+}$ (0.64 Å) in the tetragonal coordination [21]. For $FeSe_{1-\delta}$ it was also observed previously that Co doping decreases the 'a' and 'c' lattice parameters [22]. However, for $FeSe_{1-\delta}$ with Ni doping, it was previously found that the 'a' lattice parameter increases and the 'c' lattice parameter decreases [22], which is different than our finding. It was also observed in Cu doped $FeSe_{1-\delta}$ compound, the 'a' lattice parameter increases and 'c' lattice parameter decreases [17]. Fig. 3 shows the variation of lattice volume 'V', which decreases with both Co and Ni doped samples. It seems that the volume of unit cell decreases with the Co and Ni doping. Volume of Ni doped samples is relatively smaller that the Co doped.

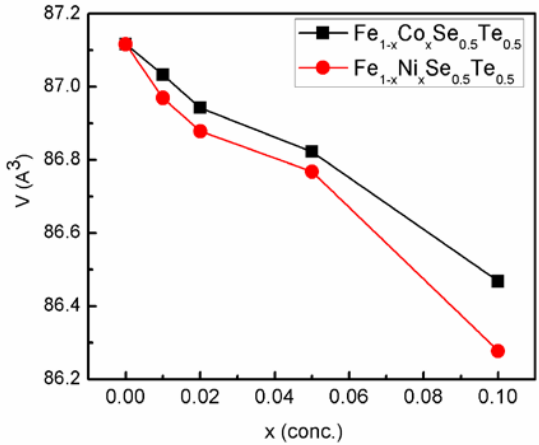

Fig. 3. Variation of lattice volume as a function of dopant concentration.

The DC magnetic susceptibility plots of $Fe_{1-x}M_xTe_{0.5}Se_{0.5}$ (M = Co, Ni and x = 0.0 0.02, 0.05) samples, are given in Figure 4. The $T_c$ for pristine compound is around 14 K, as viewed from diamagnetic onset in both the real part DC (Fig. 4) and the AC magnetic susceptibility (Fig.5). With an increase in Ni content the $T_c$ suppresses to around 4 K for 2at% doping and 5at% Ni doped sample is not superconducting. On the other hand for Co substitution although $T_c$ is decreased to 12 K and 6 K for 2at% and 5at% samples, the relative depression is much smaller than as for Ni substitution. These results are depicted in Figures 4 & 5. The large suppression of $T_c$ by Ni substitution is due to the in-plane lattice parameter for Ni doped samples is larger whereas the out-of-plane lattice parameter is smaller as compared to Co doped samples and hence large disorder in case of Ni than for Co doping. Fig. 5 shows the real and imaginary part of AC susceptibility at 33 Hz frequency and 1Oe AC drive field amplitude for the studied samples. It is also seen that superconducting transition temperature also decreases in AC susceptibility measurements with the substitution of Ni and Co

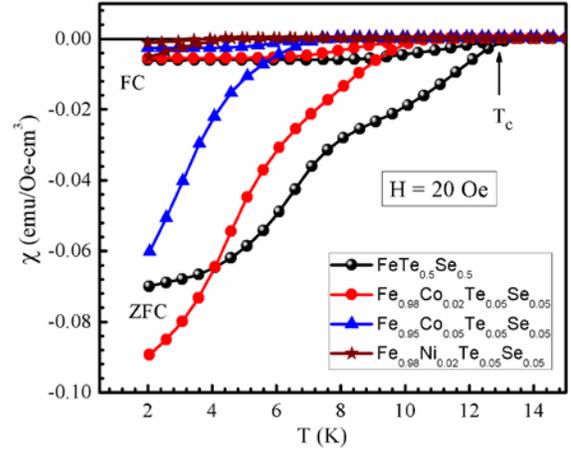

at Fe site in doped $FeSe_{0.5}Te_{0.5}$ superconductor.

Fig. 4. DC magnetic susceptibility of $FeSe_{0.5}Te_{0.5}$, $Fe_{0.98}Co_{0.02}Se_{0.5}Te_{0.5}$ and $Fe_{0.95}Co_{0.05}Se_{0.5}Te_{0.5}$ and $Fe_{0.98}Ni_{0.02}Se_{0.5}Te_{0.5}$ in ZFC and FC mode at 20 Oe.

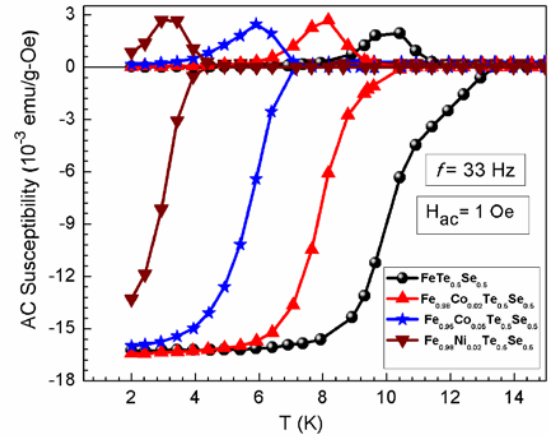

Fig. 5. Real and imaginary part of AC susceptibility for $FeSe_{0.5}Te_{0.5}$, $Fe_{0.98}Co_{0.02}Se_{0.5}Te_{0.5}$ and $Fe_{0.95}Co_{0.05}Se_{0.5}Te_{0.5}$ and $Fe_{0.98}Ni_{0.02}Se_{0.5}Te_{0.5}$ measured at frequency $f$ = 33Hz and $H_{ac}$ = 1 Oe.

## IV. CONCLUSION

It is found that there is a strong suppression of superconducting transition temperature with both 3d transition metal Co and Ni substitution. However, in-plane lattice parameter for Ni doped samples is larger whereas the out-of-plane lattice parameter is smaller as compared to Co doped samples and hence large disorder in case of Ni than for Co doping and the superconducting transition decreases fast with Ni doping as compare to Co. Summarily, our results indicate that disorder plays an important role in suppression of superconductivity in $FeSe_{0.5}Te_{0.5}$ superconductor.

## ACKNOWLEDGMENT

The authors from NPL would like to thank Prof. R. C. Budhani (Director, NPL) for his interest in the present study.

One of the authors, Anuj Kumar would also thank Council of Scientific and Industrial Research (CSIR) New Delhi, Govt. of India for providing financial support through Senior Research Fellowship (SRF) to peruse his Ph. D work.